\documentclass[aps,prl,showpacs,twocolumn,superscriptaddress]{revtex4}
\usepackage{graphicx}
\usepackage{amsmath}
\usepackage{amsfonts}
\usepackage{amssymb}
\usepackage{amsbsy}
\usepackage{ulem}
\usepackage{color}
\usepackage{times}

\begin{document}
\title{Cyclotron resonance in bilayer graphene}

\author{E. A. Henriksen}
\altaffiliation{Electronic address: erik@phys.columbia.edu}
\affiliation{Department of Physics, Columbia University, New York, New York 10027, USA}

\author{Z. Jiang}
\affiliation{Department of Physics, Columbia University, New York, New York 10027, USA}
\affiliation{National High Magnetic Field Laboratory, Tallahassee, Florida 32310, USA}

\author{L.-C. Tung}
\affiliation{National High Magnetic Field Laboratory, Tallahassee, Florida 32310, USA}

\author{M. E. Schwartz}
\affiliation{Department of Physics, Columbia University, New York, New York 10027, USA}

\author{M. Takita}
\affiliation{Department of Physics, Columbia University, New York, New York 10027, USA}

\author{Y.-J. Wang}
\affiliation{National High Magnetic Field Laboratory, Tallahassee, Florida 32310, USA}

\author{P. Kim}
\affiliation{Department of Physics, Columbia University, New York, New York 10027, USA}

\author{H. L. Stormer}
\affiliation{Department of Physics, Columbia University, New York, New York 10027, USA}
\affiliation{Department of Applied Physics and Applied Mathematics, Columbia University, New York, New York 10027, USA}
\affiliation{Bell Labs, Alcatel-Lucent, Murray Hill, New Jersey 07974, USA}

\date{\today}

\pacs{78.66.Tr; 71.70.Di; 76.40.+b.; 78.30.Na}

\begin{abstract}
We present the first measurements of cyclotron resonance of electrons and holes in bilayer graphene.  In magnetic fields up to $\textrm{B}=18$ T we observe four distinct intraband transitions in both the conduction and valence bands. The transition energies are roughly linear in B between the lowest Landau levels, whereas they follow $\sqrt{\textrm{B}}$ for the higher transitions.  This highly unusual behavior represents a change from a parabolic to a linear energy dispersion.  The density of states derived from our data generally agrees with the existing lowest order tight binding calculation for bilayer graphene.  However in comparing data to theory, a single set of fitting parameters fails to describe the experimental results.
\end{abstract} 

\maketitle

Experiments into the properties of single layer graphene have demonstrated the existence of unusual charge carriers, analogous to massless, chiral Dirac particles having a Berry's phase of $\pi$ and resulting in a new, half-integer quantum Hall effect \cite{novoselov05,zhang05}.  Bilayer graphene has shown equally exciting properties, creating a system of massive chiral particles with a Berry's phase of $2 \pi$ and exhibiting another, distinct quantum Hall effect \cite{novoselov06}.  Both phenomena derive from the unusual energy dispersion of graphene leading to the presence of a distinctive Landau level (LL) at zero energy \cite{semenoff84,haldane88,mccann06,geim07}.

The peculiar dispersion relations of graphene create unique LL spectra in the presence of a magnetic field, B.  In single layer graphene the linear dispersion leads to distinctive $\sqrt{\textrm{B}}$- and index-dependent energies for the LLs \cite{haldane88,gusynin07} which have recently been observed in infrared (IR) cyclotron resonance (CR) \cite{jiang07a,deacon07}.  In contrast, bilayer graphene exhibits a very unusual hyperbolic dispersion, which is predicted to have a more intricate index dependence and, uniquely, a {\it varying} B-field dependence.  In a nearest-neighbor, tight-binding approximation at zero magnetic field, bilayer graphene is a semimetal with degeneracies occurring at two inequivalent points at the corners of the hexagonal Brillouin zone \cite{mccann06,nilsson06}.  The low-energy part of the conduction and valence bands consists of two concentric hyperbolae, offset by the interlayer coupling energy $\gamma_1$, with energies given by 

\begin{equation} 
(2\textrm{E} \pm \gamma_1)^2 - 4 \hbar^2 \tilde{c}^2 k^2 = \gamma_1^2
\end{equation}

\noindent where $\tilde{c}$ is the band velocity and $\hbar$ is Planck's constant.  The lower energy branches meet at the charge neutrality point as depicted in the upper right inset to Fig. 1.  This zero field dispersion has been explored via angle resolved photoemission spectrometry (ARPES), demonstrating such a band arrangement \cite{ohta06}.  In a magnetic field, the lower branches are expected to quantize into LLs with energies \cite{pereira07,koshino07}

\begin{equation} 
\begin{split}
\textrm{E}_n = \frac{\textrm{sgn}(n)}{\sqrt{2}} & [ (2|n| + 1) \Delta^2 + \gamma_1^2 \\
& - \sqrt{\gamma_1^4 + 2 (2|n|+1) \Delta^2 \gamma_1^2 + \Delta^4}]^{\frac{1}{2}}.
\end{split}
\end{equation}

\noindent   Here, the index $n=\pm1, \pm2...$ describes a set of electron ($+$) and hole ($-$) LLs which are four-fold degenerate, accounting for spin and valley degrees of freedom.  The unusual zero-energy $n=0$ LL is 8-fold degenerate \cite{mccann06,novoselov06}.  The term $\Delta$ is short for $\sqrt{2 e \textrm{B} \tilde{c}^2 \hbar}$, where e is the electron charge.  With $\gamma_1$ set to zero, Eq. 2 contracts to give the LL energies for single layer graphene.

The B-field dependence of Eq. 2 differs sharply from both the LL energies of single layer graphene, and from traditional 2D systems with a parabolic dispersion for which the energies are evenly spaced and linear in B.  Instead, at low energies, $\textrm{E}_n \ll \gamma_1$, Eq. 2 approaches a linear-in-B spectrum, while at higher energies, $\textrm{E}_n \approx \gamma_1$, it approaches a $\sqrt{\textrm{B}}$-dependence \cite{abergel07,pereira07}.  Thus the LL spectrum of bilayers is expected to move from that characteristic of a parabolic dispersion to that of a linear dispersion.

Here we report CR measurements via IR spectroscopy on bilayer graphene, in magnetic fields up to $\textrm{B}=18$ T.  In both the conduction and valence bands, we resolve four LL transitions whose resonance energies decrease with increasing LL index and decreasing B-field.  The B  dependence of the resonance energies is close to $\sqrt{\textrm{B}}$ for all data, except the transitions between the lowest LLs for which the dependence is roughly linear in B.  The observed energies agree qualitatively with the predicted change in B-field dependence, but we find the changeover to a $\sqrt{\textrm{B}}$ behavior to occur at lower energies, and more suddenly, than expected.

The CR data can be exploited to arrive directly at the density of states (DOS) for bilayer graphene, finding it to be roughly linear in energy with a constant offset, as expected for a hyperbolic dispersion.  Details of this data reduction indicate contributions to the CR energies that go beyond a simple, one particle band structure picture.

The device used in this study consists of a 400 $\mu$m$^2$ flake of bilayer graphene extracted from bulk Kish graphite and deposited on a Si/SiO$_2$ substrate.  This substrate is semi-transparent in the IR yet sufficiently conductive to serve as a gate, and is thinned and wedged to suppress Fabry-Perot interference of the incident IR light.  Standard microlithography techniques are used to electrically contact the bilayer.  The sample resides on a moveable stage at the focus of a parabolic cone.   Transport properties are measured with standard lock-in techniques.  All experiments are performed in liquid helium in a magnetic field perpendicular to the graphene sheet, utilizing a Bruker IFS 66v/S Fourier-transform IR spectrometer coupled with a composite Si bolometer.  A gate voltage V$_\textrm{g}$ applied to the substrate varies the carrier density in graphene continuously from holes to electrons.  Varying V$_\textrm{g}$ amounts to shifting the Fermi level through the LLs, changing the LL filling factor $\nu$ and creating quantum Hall plateaus with R$_{xy} =h/(\nu e^2)$.  A representative trace of R$_{xy}$ taken at $\textrm{B}=18$ T is shown as the upper left inset to Fig. 1.  Clear quantum Hall plateaus appear at $\nu = \pm4, \pm8,$ and $-12$, demonstrating the 8-fold degeneracy of the $n=0$ LL and the 4-fold degeneracy for all other LLs, identifying this sample as a graphene bilayer \cite{novoselov06,mccann06}.

The IR absorption of bilayer graphene is determined as in Ref. \cite{jiang07a}.  Time-averaged IR transmission scans are taken as a function of energy at constant magnetic field for two different filling factors (V$_\textrm{g}$ values), and one is divided by the other.  As a result, a transmission minimum in the numerator (denominator) appears as a dip (peak) on a background of unity.  In all cases, we use $\nu=\pm4$, where the Fermi level lies between the $n=0$ and $n=\pm1$ LLs, as the normalizing scan for all other transitions.  Figure 1 shows three representative normalized traces at B-fields of 18 T, 14 T, and 10 T, with V$_\textrm{g}$ set so the corresponding filling factors are $\nu=-8,+12$, and $+16$ respectively.  Most sharp noise spikes are due to harmonics of 60 Hz, while the noisy band around 90 meV is due to strong IR absorption in the optical window.   For all traces, two broad resonances consisting of one peak (due to the normalizing $\nu=\pm4$ trace) and one dip indicate the CR positions.  The white traces overlaid on the IR data are best Lorentzian fits from which we determine the resonance position and  width.  The integrated intensity of the resonances  increases roughly linearly with B.  The widths of the $\nu=\pm4$ transitions are generally twice the widths of the transitions at higher filling factors.  A typical halfwidth of 14 meV for the lowest LL transitions corresponds to a carrier scattering rate of $ 2 \times 10^{13}$ s$^{-1}$ and a mobility of 3000 cm$^2$/Vs, in reasonable agreement with values from the transport data.

\begin{figure}[t]
\includegraphics[width=8.5cm]{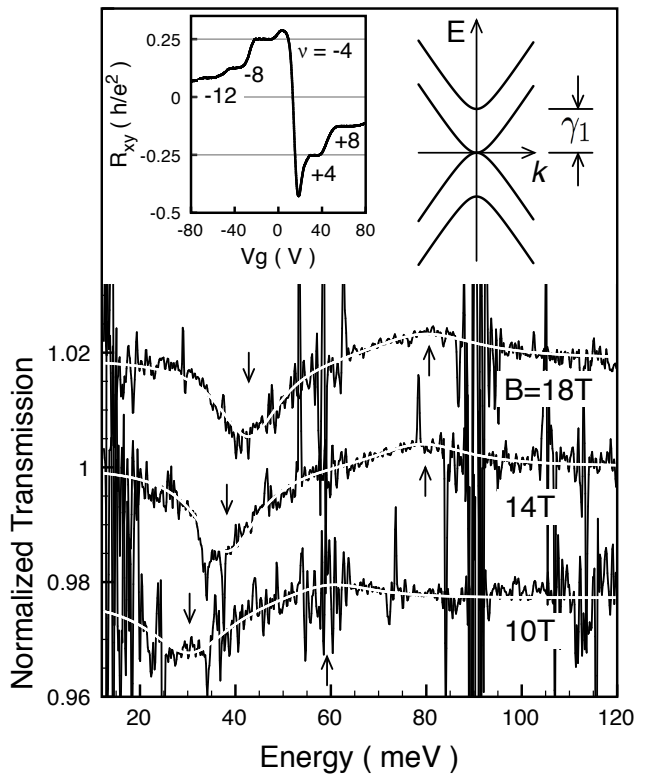}
\caption{Representative traces showing normalized IR transmission data.  In each, the resonance dip is due to the CR transition at filling factor $\nu=-8$ for $\textrm{B}=18$ T, $\nu=12$ for 14 T, and $\nu=16$ for 10 T, while the resonance peak corresponds to the CR at $\nu=\pm4$ of the normalizing scans.  White overlay traces are best Lorentzian fits, with arrows indicating dip and peak positions.  The traces are offset vertically.  Upper left inset: quantum Hall effect of bilayer graphene sample measured $\it{in~situ}$ at $\textrm{B} = 18$ T.  Upper right inset: schematic of the zero-field dispersion relation of bilayer graphene.}
\end{figure}

Our complete set of CR energies is shown in Fig. 2, comprising eight separate transitions at five B-fields.  The transitions are identified by filling factor, and plotted as a function of B.  The data can be clearly distinguished by their B-dependence.  While the data in panels (b-d) follow a $\sqrt{\textrm{B}}$ behavior (dashed lines), those in panel (a) are roughly linear in B.  Additionally, in all panels there is an apparent asymmetry between electron and hole data, with the energies of equivalent transitions differing by up to 20\%.  Such asymmetry, although on a smaller scale, has also been observed in IR data of single layer graphene \cite{deacon07,jiang08}. Its origin is presently unresolved, but is speculated to be either intrinsic \cite{deacon07} or caused by residual charged impurities \cite{tan07,jiang07b}.

\begin{figure*}[t]
\includegraphics[width=\textwidth]{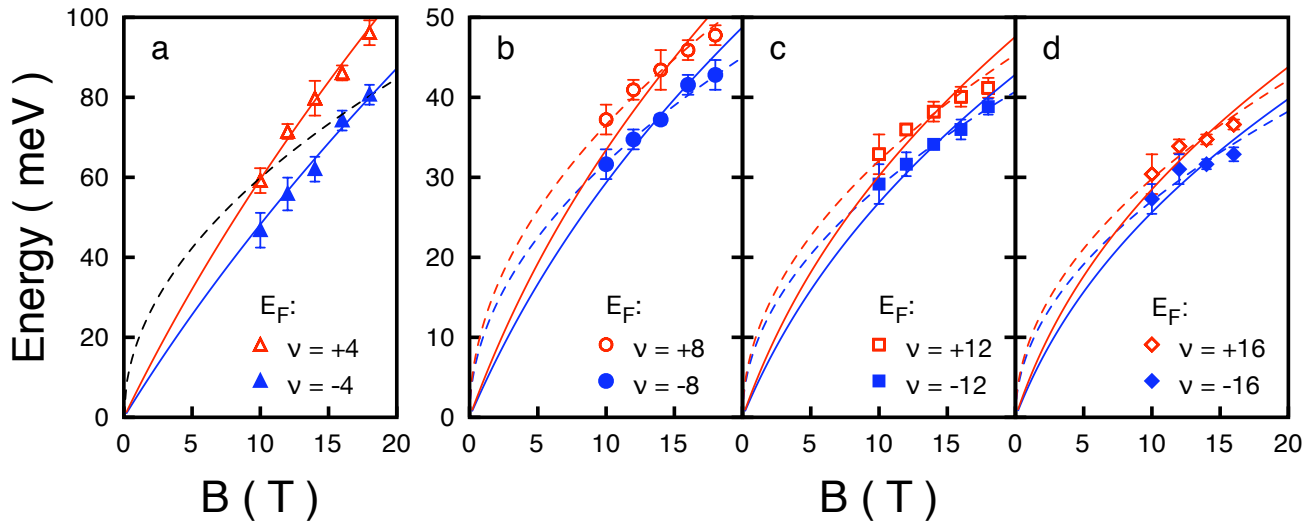}
\caption{LL transition energies as a function of B.  Open red (filled blue) symbols represent transitions between electron (hole) LLs.  Solid lines represent best fits using Eq. 2 with $\gamma_1 = 0.35$ eV and the band velocity $\tilde{c}$ as a fitting parameter.  Dashed lines are fits to Eq. 2 using $\gamma_1 = 0$.}
\end{figure*}

At first glance the data in Fig. 2 appear to match theoretical expectations: for the lowest transitions the behavior is close to linear in B, while transitions at higher energies follow $\sqrt{\textrm{B}}$.  In particular, we find a good fit of Eq. 2 to both $\nu=\pm4$ data sets in panel (a), where V$_\textrm{g}$ set to $\nu=-4$ ($+4$) corresponds to an intraband transition $n=-1 \to 0$ ($0 \to 1$).  Assuming $\gamma_1 = 0.35$ eV \cite{yan07} and leaving $\tilde{c}$ as a fitting parameter we deduce $\tilde{c} = 1.15\times10^6$ m/s and $1.02\times10^6$ m/s for electrons and holes, respectively, shown as solid lines in Fig. 2(a).  In contrast, for Fig. 2(b-d) reasonable fits can only be achieved using a value $\gamma_1 \ll 0.35$ eV.  The dashed lines in panels (b-d) are fits with $\gamma_1=0$, leading to $\textrm{E} \propto \sqrt{\textrm{B}}$, which clearly is an improvement over using $\gamma_1$ = $0.35$ eV.  This is surprising, since although the spectrum given by Eq. 2 does approach a $\sqrt{\textrm{B}}$ dependence  for energies of the order of $\gamma_1$, we see this behavior at energies  as low as $\textrm{E} \approx \gamma_1/4$.  Moreover, one set of $\gamma_1$ and $\tilde{c}$ values in Eq. 2 should describe all LL energies. Yet, for panels (b-d) it is clear that setting $\gamma_1=0$ and fitting $\tilde{c}$ over a range of values gives better fits to the data.  These fits-- which mimic the dispersion of single layer graphene--  result in $\tilde{c} =  0.6-0.7 \times 10^6$ m/s for the $\nu=\pm8$ data, rising to $0.8-0.9\times 10^6$ m/s and $0.9-1.0\times10^6$ m/s for $\nu=\pm12$ and $\pm16$.  For clear comparison, the solid lines in Fig. 2 (b-d) show fits retaining $\gamma_1=0.35$ eV, and give $\tilde{c} =  1.0-1.2\times10^6$ m/s.  In panel (a), the dashed line is a fit using $\gamma_1=0$, which for these two transitions is clearly a poor fit.  Therefore, instead of showing a smooth transition from a linear-in-B to a linear-in-$\sqrt{\textrm{B}}$ dependence with increasing energy, these data reflect a band structure that suddenly becomes linear at energies far lower than expected.

While Fig. 2 attempts to fit the parameters of a theoretical dispersion to match our results, our CR data can be presented in a fashion that directly arrives at the DOS.  The procedure is as follows (see inset to Fig. 3): the total number of states in a LL is $n_{\textrm{LL}} = d e \textrm{B} / h$, where $d$ is the LL degeneracy. This can be viewed as a condensation of the zero-field DOS in the vicinity of each LL energy into an equal number of states in the LL; half of them deriving from energies above, the other half from energies below the LL. This statement is exact for single layer graphene, while for bilayers it is exact in the high LL limit, and has no larger than 7\% error for LL $n=1$.  The zero-field DOS is then derived from
\begin{equation} 
\textrm{g}(\textrm{E}'_{n,n+1}) = \frac{d' e \textrm{B}}{h}\frac{1}{\Delta \textrm{E}_{n,n+1}},
\end{equation}
\noindent where $\Delta E_{n,n+1}$ is the measured transition energy, $n$ and $n+1$ are the initial and final LL, and $\textrm{E}'_{n,n+1}$ is the energy midway between these LLs.  The term $d' e \textrm{B} / h$ is the number of states in the zero-field DOS that lie $\it{between}$ LLs.  In bilayers, $d' \approx 4$ for all transitions except for those to/from $n=0$, where we approximate it as $d' = 6$ accounting for the 8-fold degeneracy at $\textrm{E}_0=0$ and the small error noted above.   $\textrm{E}'_{n,n+1}$ is simply determined by ``stacking'' the transition energies, $e.g. ~\textrm{E}'_{2,3} = \Delta \textrm{E}_{0,1} + \Delta \textrm{E}_{1,2} + \frac{1}{2}\Delta \textrm{E}_{2,3}$.

This method works very well for single layer graphene, using previous data for the $n=0\to1$, $-1\to0$, $2\to3$ and $-3\to-2$ LL transitions from Ref. \cite{jiang07a}, as shown in the lower trace of Fig. 3.  The so-derived DOS is linear in energy and has an intercept near the origin, reproducing closely the form of the single particle DOS, $\textrm{g(E)}=2 \textrm{E} / (\pi \hbar^2 \tilde{c}^2)$.  Fitting the data with this DOS gives a band velocity $\tilde{c} = 1.04 \times 10^6$ m/s.

The equivalent data reduction performed on our bilayer data appears in the upper left portion of Fig. 3.  Although the scatter of the data is considerably higher than for the single layer data, overall, the resulting energy dependence is linear with a constant offset, as expected from theory for a hyperbolic dispersion.  This is quite satisfactory, since no particular dispersion was assumed to generate the graphs of Fig. 3.  As in Fig. 2 we observe an electron-hole asymmetry. From least square fits to the expected DOS, $\textrm{g(E)}=( 2 \textrm{E} + \gamma_1 ) / (\pi \hbar^2 \tilde{c}^2)$, we estimate $\gamma_1=0.43$ eV and $0.52$ eV for electrons and holes, respectively.  These values are close to those previously reported on bilayer graphene in ARPES and Raman experiments \cite{ohta06,yan07}.  For both electrons and holes, the band velocity is estimated at $\tilde{c}=1.2\times10^6$ m/s.  This is higher than for single layer graphene, but includes the single layer value within error.  In fact, closer inspection of the bilayer data of Fig. 3 reveals that the scatter is not random, but correlated.

\begin{figure}[t]
\includegraphics[width=8.5cm]{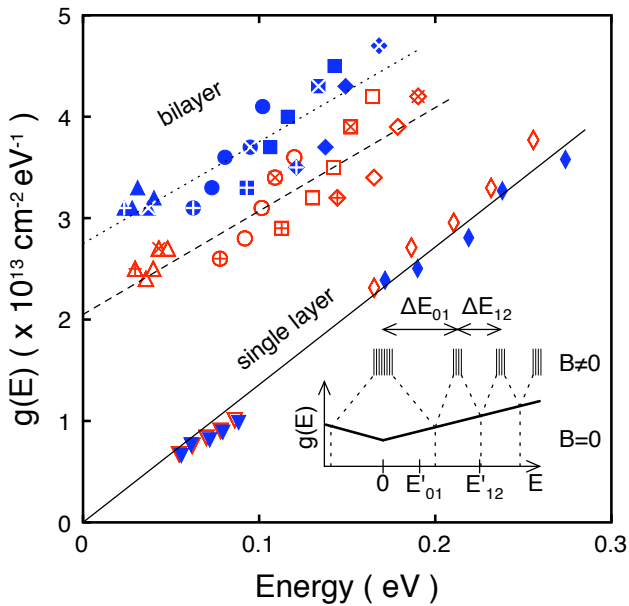}
\caption{DOS as a function of energy, derived from LL transition energies for single layer and bilayer graphene.  Open red (filled blue) symbols represent electron (hole) data.  For single layer data \cite{jiang07a}, inverted triangles are $n=-1 \to 0$ and $0\to1$ LL transitions; thin diamonds are $-3\to-2$ and $2\to3$.  For bilayer data, the symbols are the same as Fig. 2.  The solid line fit through the single layer data yields$\tilde{c} = 1.04 \times 10^6$ m/s.  Dashed (dotted) lines are fits to the bilayer electrons (holes) yielding $\gamma_1 = 0.43$ eV and $0.52$ eV, respectively.  For both, $\tilde{c} = 1.20 \times 10^6$ m/s.  Symbols overlaid with $+~(\times)$ indicate data at $\textrm{B}=10~(16)$ T.}
\end{figure}

Two trends stand out in Fig. 3.  Since the symbols used here are identical to those in Fig. 2, it is apparent that data points grouped by transition ($\nu$) form lines that are much steeper than the average slope and tend to a y-intercept ($\gamma_1/\tilde{c}^2$) much closer to the origin.  The increased slopes imply generally lower band velocities (by $\sim 30\%$) than the average.  These trends are consistent with the $\gamma_1=0$ fits in Fig. 2 (b-d), in which $\gamma_1$ and $\tilde{c}$ were derived from data grouped by transition.  On the other hand, data grouped by B field (e.g. B$=10$ T and $16$ T are marked $+$ and $\times$ in Fig. 3) show roughly the same slope as the average ($1.2 \times10^6$ m/s), but are offset in the vertical direction and extrapolate to different y-intercepts.  This implies a band velocity not far from the average, and a $\gamma_1$ that generally falls between 0.3 eV to 0.7 eV, depending on B.  The origins of these correlations in Fig. 3 are unclear.  However they are foreshadowed by our inability to fit the data of Fig. 2 with one $\gamma_1$-$\tilde{c}$ parameter set, and are therefore not the result of approximations made in arriving at Fig. 3.  In any case, in spite of these approximations, the DOS in Fig. 3 derived for single layer graphene is well behaved.  An obvious culprit is the naive interpretation of LL transition energies in terms of a one particle band structure.  LL transitions in single layer graphene have already shown inconsistencies with a one particle picture at the 5\% level, with the deviation attributed to many particle effects as large as 30\% \cite{jiang07a,iyengar07}.  Similar effects in bilayer graphene have been anticipated, as Kohn's theorem does not hold for bilayers either \cite{abergel07,kusminskiy07}.

In summary, we have performed IR absorption measurements in bilayer graphene and observed multiple cyclotron resonance features.  We find the B-field dependence of these LL transitions to be roughly linear for the lowest transition for both electrons and holes, while the remaining higher LL transitions display a clear $\sqrt{\textrm{B}}$-dependence. This highly unusual, varying B-dependence reflects qualitatively the expected behavior for the theoretically predicted hyperbolic dispersion relation.  In detail, however, the transitions do not follow such a simple picture and suggest that additional physics contributes to the transition energies.

We would like to thank Kun Yang, M. Koshino, and I.L. Aleiner for useful discussions.  This work is supported by the DOE (DE-AIO2-04ER46133, DE-FG02-05ER46215 and DE-FG02-07ER46451), the NSF under DMR-03-52738 and CHE-0117752, ONR (N000150610138), NYSTAR, the Keck Foundation, and the Microsoft Project-Q.  The infrared measurements in this work were performed at the National High Magnetic Field Laboratory, which is supported by NSF Cooperative Agreement No. DMR-0084173, by the State of Florida, and by the DOE. We thank J. Jaroszynski and E.C. Palm for experimental assistance.  E. A. Henriksen and Z. Jiang contributed equally to this work.

\end{document}